# Methods for adjusting for covariate measurement error in flexible modelling of functional form: designing a blinded, controlled neutral comparison simulation study

Anne C.M. Thiébaut[*,1], Aris Perperouglou[2], Mohammed Sedki[1], Steve Ferreira Guerra[3], Paul Gustafson[4], Frank E. Harrell, Jr[5], Willi Sauerbrei[6], Michal Abrahamowicz[3], Laurence S. Freedman[7]; on behalf of Selection of variables and functional forms in multivariable analysis Topic Group (TG2) and Measurement Error and Misclassification Topic Group (TG4) of the international STRengthening Analytical Thinking for Observational Studies (STRATOS) Initiative

Affiliations:

1. Université Paris-Saclay, UVSQ, Inserm, CESP, Villejuif, France
2. Statistics and Data Science Innovation Hub, GlaxoSmithKline, London, UK
3. Department of Epidemiology, Biostatistics and Occupational Health, McGill University, Montreal, Canada
4. Department of Statistics, University of British Columbia, Vancouver, Canada
5. Department of Biostatistics, Vanderbilt University School of Medicine, Nashville, Tennessee, USA
6. Institute of Medical Biometry and Statistics, Faculty of Medicine and Medical Center - University of Freiburg, Freiburg, Germany
7. Information Management Services Inc., Calverton, Maryland, USA

[*]Corresponding author: e-mail: anne.thiebaut@inserm.fr, Phone: +33-1-45-59-52-66

## ABSTRACT

This article describes the design of a neutral comparison study in the context of empirical studies where the interest is in learning the functional relationship between a continuous error-prone exposure variable and a binary outcome. The performance of combinations of measurement error correction methods and flexible regression modeling techniques was compared using a simulation study. The project involved four independent teams, one devoted to data generation and evaluation, the other three to specific methods of measurement error correction (Simulation-Extrapolation, Regression-Calibration and Multiple imputation, Bayesian method). The study was conducted in three successive stages. In Stage 1, the first team simulated five datasets differing only by the true exposure-outcome functional form and distribution of true exposure. Furthermore, the implementation of flexible modeling methods (B-splines, P-splines, and fractional polynomials) was standardized. The three methods teams, blinded to the underlying data generation process, created the codes to implement their methods, and provided their results to the first team who evaluated them. These codes were then used by this team in the next Stages of the project. In Stage 2, the team simulated 150 additional datasets where other design parameters varied while using the same five exposure-outcome functions. Stage 3 consisted of simulating independent replications of each of the 150 scenarios considered in Stage 2 to quantify the sampling variance of the estimates. This work emphasizes the relevance of neutral comparison studies to fairly evaluate statistical methods aimed at addressing a complex analytical challenge, and demonstrates their feasibility through a large collaborative project.

## 1. INTRODUCTION

The STRengthening Analytical Thinking for Observational Studies (STRATOS) initiative aims to provide guidance on which statistical methods could be used for various types of observational studies (Sauerbrei *et al.*, 2014, https://stratos-initiative.org). This guidance requires reliable evidence on the comparative advantages of different competing methods. Often this evidence is provided by simulation studies (Boulesteix *et al.*, 2020). Yet, biostatisticians teach clinicians the importance of unbiased, controlled evaluation of treatments, but rarely subject their own methods to the same standards (Boulesteix *et al.*, 2017; Pawel *et al.*, 2024; Siepe *et al.*, 2025). Neutral comparison studies are those that are designed to provide such an unbiased, controlled comparison of competing statistical methods (Boulesteix *et al.*, 2024). They are especially useful for comparing existing, already described, methods, rather than a new prototype method (Boulesteix *et al.*, 2024; Heinze *et al.*, 2024). When the study design is non-blinded, the researchers involved should be neutral, in the sense that they are reasonably familiar with all compared methods and without strong prior preference for or strong aversion to one of the comparator methods (Boulesteix *et al.*, 2013).

This article addresses a complex issue where a neutral comparison study is desirable. We consider empirical studies where the primary interest is in learning the functional relationship between a continuous predictor, or exposure variable, and an outcome variable when the exposure variable is measured with error. When the shape of the exposure-outcome relationship is nonlinear, it is well known that random error in exposure measurements can mask the features of the relationship so that the observed shape may be distorted from the true shape (Carroll *et al.*, 2006). For example, under specific circumstances (a single error-prone covariate with classical measurement error), the estimated relationship is likely to look more linear than it truly is (Keogh *et al.*, 2012; Ferreira Guerra *et al.*, 2026).

Various statistical methods are available for tackling this problem, with many flexible regression methods available for modelling non-linear relationships on the one hand (*e.g.*, Perperoglou *et al.*, 2019; Royston *et al.*, 1999), and measurement error correction methods on the other (Keogh *et al.*, 2020; Shaw *et al.*, 2020). However, relatively little is known about their relative performance when these two approaches are combined. To ensure a fair comparison of alternative strategies combining measurement error correction methods with flexible modeling methods, we thought it important to design a neutral comparison simulation study where researchers involved in proposing different analytical strategies would be blinded to the actual shape of the exposure-disease association. This article describes how such a neutral simulation study was designed. This was a joint work of two topic groups (TG) of the STRATOS Initiative: TG2 (Selection of Variables and Functional Forms) with expertise in the choice of functional forms for continuous variables in multivariable explanatory models, and TG4 (Measurement Error and Misclassification) with expertise in statistical methods to account for measurement error.

## 2. METHODS

The Methods section is divided into two subsections, one detailing the project organization and how we ensured a neutral comparison design, the other briefly describing the simulation study.

### 2.1 Neutral comparison design

The project involved four independent teams: one "Data Generation and Evaluation" team, and three "Methods" teams (see Appendix for the list of members in each team). Each of the latter considered different measurement error correction methods (regression-calibration and multiple imputation, simulation-extrapolation [SIMEX], and Bayesian). The teams were independent in the sense that (*i*) membership of the teams was non-overlapping, and (*ii*) the Methods teams did not confer with each other when implementing their methods. The study was conducted in successive stages (Figure 1):

Stage 1 aimed at implementing the methods under blinded conditions with respect to the data characteristics. The Data Generation and Evaluation team was in charge of simulating the data and standardizing the choice and implementation of alternative flexible modeling methods (B-splines, P-splines, and fractional polynomials [FPs]). Five datasets were generated differing only by the true exposure-disease functional form and distribution of true exposure. The Data Generation and Evaluation team shared the five datasets with the Methods teams without revealing the underlying data generation process. Each of the three Methods teams in turn was in charge of creating the codes to implement their methods. After writing their code and applying it to the five datasets, they provided their results to the Data Generation and Evaluation team who evaluated them. This evaluation was presented to the Methods teams in a blinded manner to give them feedback and motivate their interest in continuing the investigation. The alternative analytical strategies were given a blinded identification code and then ranked according to their relative overall performance, as well as separately for each of the five simulated datasets. When presenting the results, the Data Generation and Evaluation team did not reveal the identities of the analytic strategies nor the true functional form underlying each dataset. After the blinded results were presented, the Methods teams were asked to finalize their codes and submit them to the Data Generation and Evaluation team. Some of the Methods teams created different versions of their methods, and were allowed to try them out on the first datasets. At the conclusion of Stage 1, they were asked to decide finally which version they wanted to run with. The codes provided were 'generic', *i.e.*, allowed implementation of the respective methods for different datasets prepared according to the same principles. These generic codes were then used to analyze additional data simulated in the next Stages of the project.

Stage 2 aimed at evaluating the performances of the methods under a much wider range of scenarios. Using the same five functions defining the exposure-outcome relationship as in Stage 1, the Data Generation and Evaluation team simulated 150 additional datasets where the combination of four other design parameters (main study sample size, replication study size, and the measurement error variance and its distribution) were varied. After the data were generated, the Data Generation and Evaluation team ran the codes provided by the Methods team on these new simulated datasets, and evaluated the results. They then presented the evaluation results to the Methods team, this time unblinded. Only point estimates were evaluated at this stage.

Stage 3 aimed at assessing sampling variance for the methods' average performance across repeated sampling. It consisted of simulating 10 replications of each of the 150 scenarios considered in Stage 2. A further flexible modeling method, natural splines, was added at this stage in an unblinded fashion, due to concerns from members of the Methods teams that the B-spline method chosen initially may not have been optimal.

### 2.2 Simulation study

Following the recommendation of Morris *et al*. (2019), the description of the simulation study is organized according to the ADEMP structured approach: Aims, Data-generating mechanisms, Estimands, Methods, and Performance measures. The purpose of this presentation is to give an overview of the simulation study that was conducted to answer the methodological question that motivated a neutral comparison design. More details about the methods and results may be found in a separate manuscript.

### 2.2.1 Aims

The aim of the simulation study was to compare the performance of combinations of measurement error correction methods and flexible regression methods in retrieving the true functional relationship between a single continuous exposure variable and a binary outcome variable when the exposure variable is measured with classical error.

### 2.2.2 Data generation

Each dataset mimicked a nested case-control study with a 1:4 case:control ratio. A binary outcome $Y$ was generated from a logistic regression on continuous exposure $X$, *i.e.*, $\text{logit}(P(Y = 1|X)) = f(X)$ where $f(X)$ represents a possibly nonlinear exposure-outcome relationship.

We considered five scenarios inspired by examples from real health studies. For each scenario, we searched the literature for data to define plausible parameters for the functional form $f(X)$ and the distribution of true exposure $X$ (Table 1, Figure 2). To ensure that the overall event rate equals 20%, consistent with the pre-specified 1:4 case:control ratio, first a large number of pairs of $(Y, X)$ were generated at random according to the true distribution of $X$ and the relationship $\text{logit}(P(Y{=}1)) = f(X)$, as specified in Table 1. Then from this large number of pairs, $n/5$ pairs were drawn at random from the pairs with $Y{=}1$, and $4n/5$ pairs were drawn at random from the pairs with $Y{=}0$. All simulations considered only $f(X)$ un-adjusted for any other covariate.

In the first scenario, $f(X)$ was J-shaped borrowing from the relationship between body mass index (BMI) and all-cause mortality. The parameters of the functional form $f(X)$ were derived from a dose-response meta-analysis of 230 cohort studies (Aune *et al.*, 2016). The distribution of BMI as true exposure $X$ was taken as lognormal based on Silverman and Lipscombe (2022).

The second scenario assumed a linear relationship as reported in the dietary fat and breast cancer analysis from the National Institutes of Health-AARP cohort study (Thiébaut *et al.*, 2007). The exposure was expressed as log percent energy from total fat, *i.e.*, the logarithm of calories from usual dietary fat divided by energy intake in calories. Its true distribution was derived using information on the measurement error structure inferred from the biomarker-based Observing Protein and Energy Nutrition (OPEN) study (Kipnis *et al.*, 2003).

The third and fourth simulation scenarios were based on the relationship between air pollution levels and mortality rates (Cakmak *et al.*, 1999). Exposure $X$ was taken as air pollution level measured in $\mu g/m^2$. The authors postulated a threshold model for mortality rate with a linear increase above a certain threshold exposure level, denoted by $\tau$ (our third scenario). We alternatively assumed an exponential increase above the threshold exposure (our fourth scenario). Moreover, we changed the changepoint $\tau$, from below (third scenario) to above (fourth scenario) the median of $X$ because previous simulation work had shown that the position of the threshold may qualitatively impact the biases caused by measurement error (Küchenhoff and Carroll, 1997).

The fifth scenario was inspired by the Hill equation that is frequently used in pharmacological modeling to describe a saturation effect (Goutelle *et al.*, 2008). We postulated a side effect whose risk is dependent on the dose of drug $X$ that is taken by the patient, resulting in a monotonic increasing relationship that eventually reaches a plateau. The distribution of the doses was assumed to be uniform, reflecting different practices among prescribing physicians.

For each of the above scenarios, after simulating true exposure values $X$ and corresponding values of outcome $Y$, an error-prone $X^* = X + e$ was generated, where measurement error $e$ was assumed nondifferential with respect to $Y$, classical (purely random), with a normal distribution of mean 0 and variance equal to half that of the variance of true exposure $X$. In Stage 1, the main study data consisted of 15,000 individuals each with an independent realization of $(Y, X^*)$. In addition, a replication sub-study consisting of a random subgroup of 250 participants of the main study, with a second independent realization of $X^*$, was simulated. Methods teams were not informed of the theoretical measurement error variance, but they were able to estimate it from the replication data provided to them. Relatively large sample sizes were chosen in Stage 1 to ensure convergence of flexible modelling methods. Smaller sample sizes were investigated later in Stages 2 and 3.

### 2.2.3 Estimands

The estimands of interest were the values of the adjusted true function $f^*(X)$ at the pre-selected series of $X$ values. The adjustment was made to reflect the 1:4 nested case control ratio in the data generated, and was given by $f^*(X) = f(X) + \log(0.25\ (1-p) / p)$, where $p = P(Y=1)$ in the population. For each dataset, the Data Generation and Evaluation team provided the Methods teams with a sequence of 200 $X$ values, *e.g.*, for dataset 1, from 10.0 to 50.0 inclusive, in steps of 0.2 (see Table 1). The selected $X$ values covered the range of the erroneous $X^*$ values and thus were much wider than the true range of $X$. The Methods teams were expected to provide corresponding point estimates of $f^*(X)$ for each of these 200 $X$ values.

### 2.2.4 Methods

The general principles of the concurrent methods will be presented here. More details and further references may be found in (Keogh *et al.*, 2020; Shaw *et al.*, 2020).

- Flexible modeling methods

In Stage 1, three flexible modeling methods were considered. The first one was cubic B-splines with one interior knot at the median of observed $X^*$, while the boundary knots were placed at its minimum and maximum (4 degrees of freedom). The second one was P-splines with 10 interior knots, with penalty to be optimized by the Methods teams. The third one was FPs of second degree (4 degrees of freedom), that permits estimating non-monotone functions, with the same initial set of powers, but leaving it up to the Methods teams to select the optimal two powers.

After disclosure of the true models underlying the simulated data, it was decided to additionally implement natural splines as a fourth flexible modeling method in Stage 3. The natural splines (2 degrees of freedom) were specified as having three knots, one at the median, and two at the boundaries, *i.e.*, the $5^{th}$ and $95^{th}$ percentiles of $X^*$.

- Measurement error correction methods

After discussion among the project group and with other experts on measurement error adjustment methods, four methods were identified as being accepted, well-researched, and easily programmable: regression-calibration, multiple imputation, SIMEX, and Bayesian. The Methods teams were formed based on the expertise of their members with these methods.

The regression-calibration approach consists of estimating the conditional expectation of the function $f(X)$ given the error-prone covariate $X^*$ and substituting it for the true covariate in the logistic regression (Rosner *et al.*, 1989). In the multiple imputation approach, the imputed $f(X)$ consists of its conditional expectation given $X^*$ and $Y$ plus imputed values of the regression residual (Cole *et al.*, 2006). Imputation was done 10 times using different model parameter values from the corresponding estimated distributions.

The SIMEX approach is a two-step method that has been adapted to various measurement error problems (Cook and Stefanski, 1994; Carroll *et al.*, 2006). The general idea is to sequentially simulate new values of error-prone exposures with gradually increasing measurement error, and then, for each of the assumed measurement error levels, estimate the parameter of interest using the generated variables. The estimates being increasingly biased, this establishes a relationship between the estimate (and thus, implicitly, the amount of bias) and amount of added measurement error. The second step consists in extrapolating this relationship back to the case of no error. For this project, the team considered two alternative SIMEX approaches applied either (*i*) to the individual points on the curve, or to (*ii*) to the estimated B-spline or FP coefficients (the latter approach was not feasible for P-splines).

In the Bayesian approach (Gustafson, 2003), the team specified three models: an outcome model for $Y$ given $X$, an exposure model for $X$, and a measurement error model for $X^*$ given $X$, along with prior distributions for relevant parameters in each of the three sub-models. This defined a joint posterior distribution of all parameters and latent $X$ values, given all the observed data. The team used a particular Monte Carlo Markov Chain algorithm, Hamiltonian Monte Carlo (Betancourt and Girolami, 2013), as implemented in the Stan software (Stan Development Team, 2025) to obtain a Monte Carlo approximation to this posterior distribution. Two different summaries for inference were considered: (*i*) risk, corresponding to the logit of the posterior mean of risk on the $P(Y=1)$ scale, and (*ii*) logit, corresponding to the posterior mean on the $\text{logit}(P(Y=1))$ scale.

#### 2.2.5 Performance measures

The main evaluation criterion in Stage 1 was the mean squared error (MSE), that is the mean of the squared difference between the estimated $f^*(X)$, obtained using a given analytic method, and the true $f^*(X)$ over the 200 pre-selected values of $X$ that lay within limit points. These limit points were chosen by the Data Generation and Evaluation team and kept undisclosed to the Methods teams. They were defined as the 2.5$^{\text{th}}$ and 97.5$^{\text{th}}$ percentiles of the distribution of true exposure $X$, to avoid an excessive impact of estimates in the tails of the distribution. Values among the 200 estimates of $f^*(X)$ provided by the Methods teams for $X$ outside of these limits were discarded. The overall performance was evaluated as the mean of the MSEs over the scenarios. In Stage 2 and 3, the main criterion was switched to log MSE due to the right skewed distribution of MSE values.

For Stage 1 results, two sorts of benchmarks were obtained based, respectively, on either using true exposure $X$ (as a lower bound for MSE) or uncorrected B-spline and P-spline estimation applied to observed exposure $X^*$ (expected to mark the upper bound on errors). The latter was not performed for FPs, or with the extended scenarios and replications in Stages 2 and 3.

## 3. RESULTS

Here we briefly report on the Stage 1 results to emphasize the interest of the neutral comparison design. Results from the extensive simulation study, including variants of the original five scenarios and the addition of natural splines as a fourth flexible modeling method, will be presented in a separate manuscript.

Table 2 displays the unblinded results with benchmarks. It should be noted that these results were initially presented in a blinded fashion to the Methods teams, *i.e.*, were displayed in decreasing order of overall performance but with the blinded identification code replacing the labels of specific methods.

Both versions of SIMEX performed best, with a slight advantage of SIMEX applied to individual points over that applied to coefficients. Multiple imputation performed second best. The Bayesian methods and the regression-calibration methods performed, on average, more poorly than the benchmark applied to observed $X^*$. The ranking across measurement error correction methods did not vary across the flexible modeling methods except for one notable exception: the Bayesian approach combined with B-splines performed much poorer than that combined with P-splines or FPs, and worse than any of the other methods.

For some scenarios, the spline estimates obtained using uncorrected $X^*$ values had MSEs that were lower than several of the correction methods (Table 2). This, rather unexpected, finding may partly reflect the bias-variance tradeoff although measurement error correction methods decrease bias, they also increase variance, so that MSEs are not guaranteed to be lower than uncorrected analyses.

## 4. DISCUSSION

In this paper, we report a multi-center experiment of a neutral comparison study addressing a complex issue often encountered in epidemiological studies: how to retrieve the shape of an exposure-outcome association when the exposure is measured with error? This paper is not intended to provide guidance on the most appropriate analytic methods to use, which will be addressed in a separate manuscript, but rather to describe the study design for comparing such methods. Designing a neutral comparison study for addressing our statistical question was crucial. Had the Methods teams known the shapes of the relationships, they could have been tempted to adjust tuning, which would have introduced overoptimism and, thus, compromise the practical usefulness of the results.

Neutral comparison studies have been advocated to provide an unbiased, controlled comparison of competing statistical methods (Boulesteix *et al.*, 2024). In the present experiment, neutrality and maintenance of study design were ensured by the blinded design, by the participation of motivated independent international teams, and the strong commitment of a chair person. Blind analyses have notably been proposed as a correction for confirmation bias, whereby one concurrent method could be favored over the others at the outset (MacCoun and Perlmutter, 2017). Among several possible techniques, our choice was to keep the data generation process hidden to those who then implemented the concurrent methods (Dutilh *et al.*, 2019). To prevent peeking, the members of the Methods teams were from different institutions than those of the Data Generation and Evaluation team. To enforce motivation, regular virtual meetings were held for the full project group, within each Methods team, as well as within the Data Generation and Evaluation team. Similarly, Geroldinger *et al.* (2024) built upon an existing consortium of researchers willing "to learn from each other, and to openly discuss advantages and drawbacks of the respective methods". Our experiment is another illustration of an "adversarial collaboration" (Cowan *et al.*, 2020) in the sense that proponents of different measurement error correction methods have agreed and worked together to test their intuition. Combining adversarial collaboration and blind analysis is an innovative feature that has hardly ever been applied in methodological simulation research (Kreutz *et al.*, 2020).

Although Stage 1 results are clearly insufficient to draw conclusions, we should admit that they looked intriguing to the Data Generation and Evaluation team. From theoretical considerations, we had hypothesized that the Bayesian and multiple imputation methods would be expected to perform better than regression-calibration, which, in turn, would perform better than SIMEX. Instead, Stage 1 results suggested that, on average, SIMEX performed best, and multiple imputation was next best. The Bayesian method with FPs or P-splines came third, before regression-calibration, while the Bayesian method with B-splines performed poorly. In the face of such surprising results, the blinded design helped to insure against risks of conscious or unconscious biases (MacCoun and Perlmutter, 2015).

Beyond the neutral comparison design, another strength of our work is the choice of realistic functional shapes using parameters derived from real datasets (Sauer *et al.*, 2025). The choice of the five simulation scenarios was driven by the need of a variety of plausible functional relationships, and was not motivated at all by the idea that some methods may fail or better succeed in specific cases. The only exception was the linear relationship where we expected all methods to perform best but were also interested to see if measurement error may induce some spurious non-linearities. Choosing these scenarios independently from the methods implementation and in a blinded fashion ensured prevention of selective reporting and "cherry-picking" of more favorable results, as did evaluation of results by an independent team (Pawel *et al.*, 2024; Siepe *et al.*, 2024). Averaging performances across scenarios in Stage 1 provided a general idea allowing to discard variants of methods that would perform especially poorly in all instances. In Stage 2, when we considered variations of the five simulation scenarios, some were indeed motivated by the intuition that one specific method may perform more poorly (Strobl et al., 2024). Nonetheless, all methods were subjected to the same scenario variants so we believe the comparison remains fair and systematic.

A requirement for neutral comparison studies is the choice of the evaluation criteria in a rational way (Boulesteix *et al*., 2013). The choice of the performance measure was highly debated in the early phases of the project. We considered unweighted MSE and MSE weighted by the density of $X$, and mean absolute error, on either the log odds scale or the risk scale. For Stage 1, after finding that weighted and unweighted MSE were similar, we chose to present unweighted MSE on the log odds scale. For the later stages, however, we finally switched to log MSE. Other criteria would have been possible, as highlighted in Ullmann *et al.* (2025)'s categorization of performance measures for evaluating estimated associations between an outcome and continuous predictors.

We note some practical disadvantages of the blinded, controlled design. It was more difficult to organize and offered less flexibility to change or introduce new methods. This happened during the present experiment with the later introduction of natural splines that had to be done unblinded. In fact, after seeing unblinded results, members of the Methods teams expressed concerns that the B-spline method was not optimal. For this reason, a fourth method, natural splines, was considered in Stage 3. The fact that the choice was made *a posteriori* yet weakens any conclusion that can be drawn about the possible superiority of natural splines over other flexible regression methods.

Although efforts have been made to standardize the flexible modeling methods as much as possible, we cannot rule out the possibility that performance discrepancies between two measurement error correction methods using the same flexible modeling approach may result in part from some variations regarding how different Methods teams implemented the same flexible method (*e.g.*, choice of the penalty for P-splines, or the algorithm to choose best-fitting second-degree FPs). Indeed, methods teams sometimes encountered technical difficulties in combining a specific measurement error correction method with a flexible modeling approach. For example, the SIMEX coefficient variant was not feasible for P-splines.

Our study could also be considered limited by the fact that each of the competing methods was implemented by only one of the three non-overlapping team. An ideal framework would have involved several teams devoted to the same competing method. It has been recognized that several teams using the same method may indeed provide different results, possibly as a consequence of different levels of expertise (Nieβl *et al.*, 2024).

In conclusion, this work emphasizes the relevance of neutral comparison studies to fairly evaluate statistical methods aimed at addressing a complex analytical challenge, and demonstrates their feasibility through a large collaborative project. Blinded, team-based, neutral comparison studies seem most appropriate when knowing the features of the data could influence the tuning of the methods whereas truth is unobservable in real practice. They are also advisable when one of the methods can currently be conducted only by the team who developed the method.

**FIGURES**

**Figure 1** Schematic representation of the project organization

**Figure 2** Five forms of $f(X)$ used in the simulations: (1) J-shaped, (2) linear, (3) threshold below the median and linear increase, (4) threshold above the median and exponential increase, (5) saturation. The red dotted vertical bars correspond to the limit points

**Acknowledgements**

The authors are grateful to all members of Measurement Error and Misclassification Topic Group (TG4) of the international STRengthening Analytical Thinking for Observational Studies (STRATOS) initiative for providing valuable comments on this work, in particular (in alphabetical order) Drs. Sabine Hoffmann and Pamela Shaw. They would also like to thank Georg Heinze (STRATOS TG2) and an anonymous referee for constructive and helpful comments that improved the manuscript.

**Conflict of Interest**

The authors have declared no conflict of interest.

**Appendix**

**A.1 List of members of each team (in alphabetical order)**

- Data Generation and Evaluation team

Anne Thiébaut (TG4), Laurence Freedman (TG4), Aris Perperoglou (TG2), and Mohammed Sedki (affiliate)

- Simulation-Extrapolation (SIMEX) team

Michal Abrahamowicz (TG2) and Steve Ferreira Guerra (affiliate)

- Regression-calibration & Multiple imputation team

Victor Kipnis (TG4), Brian Barrett (affiliate), Matthew Chaloux (affiliate), Kevin Dodd (TG4), Douglas Midthune (TG4), and Amer Moosa (affiliate)

- Bayesian team

Paul Gustafson (TG4), Raymond Carroll (TG4), Frank Harrell (TG2), and Nadja Klein (affiliate)